\documentclass[aps,prb,twocolumn,showpacs,superscriptaddress]{revtex4}
\usepackage{graphicx}
\usepackage{bm}
\begin{document}
\newcommand{\beq}{\begin{equation}}
\newcommand{\eeq}{\end{equation}}

\title{\boldmath Merging of single-particle levels in finite Fermi systems}
\author{V.~A.~Khodel}
\affiliation{ Russian Research Centre Kurchatov
Institute, Moscow, 123182, Russia}
\affiliation{ McDonnell Center for the Space Sciences and
Department of Physics, Washington University,
St.~Louis, MO 63130, USA }
\author{J.~W.~Clark}
\affiliation{ McDonnell Center for the Space Sciences and
Department of Physics, Washington University,
St.~Louis, MO 63130, USA }
\author{Haochen Li}
\affiliation{ McDonnell Center for the Space Sciences and
Department of Physics, Washington University,
St.~Louis, MO 63130, USA }
\author{M.~V.~Zverev}
\affiliation{ Russian Research Centre Kurchatov
Institute, Moscow, 123182, Russia}
\date{\today}

\begin{abstract}
Properties of the distribution of single-particle levels adjacent
to the Fermi surface in finite Fermi systems are studied, focusing
on the case in which these levels are degenerate.  The
interaction of the quasiparticles occupying these levels lifts
the degeneracy and affects the distance between the closest
levels on opposite sides of the Fermi surface, as the number
of particles in the system is varied.  In addition to the familiar
scenario of level crossing, a new phenomenon is uncovered,
in which the merging of single-particle levels results in
the disappearance of well-defined single-particle excitations.
Implications of this finding are discussed for nuclear, solid-state,
and atomic systems.

\end{abstract}

\maketitle
Advanced technologies seek to exploit the properties of objects
of nanometer size in the design of new materials and devices.
The behavior of the electronic system within such a nanoscale
object is largely determined by the structure of individual
single-particle (sp) levels.  It therefore seems opportune
to revive the fundamental study of single-particle aspects of
finite Fermi systems as developed many years ago for atomic
nuclei,\cite{migdal} in the expectation that the findings may
also be of value for electronic systems of current technological
importance.

In homogeneous matter, all the relevant measurable quantities,
such as various susceptibilities, are functions of a single
momentum transfer variable $q$.  Inhomogeneous systems
with a uniform distribution of sp levels
possess basically the same properties.  However, the situation
changes when the spectrum of their sp excitations is degenerate.
This degeneracy implies the existence of a new energy scale
$D_{\rm min}$, given by the difference between the energies of
the closest sp levels lying on opposite sides of the Fermi
surface.  The properties of such systems exhibit striking departures
from what is found in homogeneous matter. To offer a prominent
example, consider the ground-state energy $E_0(A)$ of atomic
nuclei as a function of mass number $A$.  For most nuclei,
this quantity is well described by the Bethe-Weisz\"acker
liquid-drop formula.  However, nuclei with a magic number of
protons or neutrons have spherical form, and the relevant energy
scale $D_{\rm min}$ is several times larger than the average
distance between neighboring sp levels in non-magic nuclei.
This energy spacing provides a shell correction $\delta E_s$,
lowering the liquid-drop binding energy and rendering the
ground states of known magic nuclei stable with respect
to any mode of decay.\cite{bohr} Another example is associated
with the degeneracy of the sp spectrum of the two-dimensional
electron gas in an external magnetic field.  In this case,
the degeneracy gives rise to a step-like behavior of the
chemical potential $\mu(A)$, triggering oscillations of
thermodynamic properties.\cite{shoenberg}

Customary explanations of such extraordinary behavior do not take
account of the alteration of key quantities due to interactions
between added particles under variation of their number. In many
cases such an approximation is justified, since these interactions
do not affect the deviations mentioned above, even if the sp
levels cross one another.
However, we shall demonstrate that the dependence of sp energies
$\epsilon_s=E_s(A+1)-E_0(A)$ on quasiparticle occupation numbers
$n$, inherent in  Fermi-liquid (FL) theory,
allows for an alternative scenario.  The familiar crossing
of sp levels is replaced by a merging of these levels,
a new phenomenon that leads to the disappearance of well-defined sp
excitations and drastic departures from predictions of standard
FL theory.

To gain insight into this unconventional scenario, we first
consider a schematic model involving three neutron levels
in an open shell of a spherical nucleus.  The levels are denoted
$-$, $0$, and $+$, in order of increasing energy, and the
distance between $-$ and $0$ and between $0$ and $+$ has
the same value $D$.  As usual, the sp energies and wave functions
$\varphi_{\lambda}({\bf r})=R_{nl}(r)\Phi_{jlm}({\bf n})$ as well,
are solutions of equation
$ [p^2/2M+\Sigma({\bf r},{\bf p})]\varphi_{\lambda}({\bf r}) =
\epsilon_{\lambda}\varphi_{\lambda}({\bf r})$, where
$\Sigma$ stands for the self-energy. In even-even spherical
nuclei, which have total angular momentum $J=0$ due to pairing
correlations, the energies $\epsilon_{\lambda}$ are
independent of the magnetic quantum number $m$ associated
with the total sp angular momentum $j$.  We suppose that
the level $-$ is filled, the level $+$ is empty, and $N$
neutrons are added to the level $0$, changing the density by
$\delta\rho(r)=NR^2_{n_0l_0}(r)/4\pi$. We assume that
$l_-\neq l_0\neq l_+\sim A^{1/3}\gg 1$.

It is our goal here to explore the consequences of the dependence
of the sp energies $\epsilon_{\lambda}(n)$ on the distribution $n$.
In what follows, we shall retain only a major, spin- and
momentum-independent part of the self-energy $\Sigma$ and
a primary, $\delta(r)$-like portion of the Landau-Migdal
interaction function $f$.  Accordingly, the FL relation between
$\Sigma $ and the density $\rho$ responsible for the variation
of $\epsilon_{\lambda}(n)$ with $n$ is simplified to\cite{migdal}
$\delta\Sigma(r)=f[\rho(r)]\delta \rho({\bf r})$.
For the sake of simplicity, the diagonal and nondiagonal
matrix elements of $f$ are assigned the respective values
\begin{eqnarray}
u&=&\int R_{nl}^2(r)f\left[\rho(r)\right]R^2_{nl}(r)r^2dr/4\pi \,, \\
\label{mel1}
w&=& \int R_{nl}^2(r)f\left[\rho(r)\right]R^2_{n_1l_1}(r)r^2dr/ 4\pi  \,,
\label{mel2}
\end{eqnarray}
independently of the quantum numbers $nl,\,n_1l_1$.

A simple estimate of the ratio $u/w$ is obtained using a
semiclassical approximation $R_{nl}(r)\sim  r^{-1}\cos\int p(r)dr$,
with the result $u\simeq 3w/2$.
We next observe that $f(\rho)$ is positive at densities
close to equilibrium,\cite{migdal} but changes sign as
$\rho \to 0$; hence the signs of $u$ and $w$ may
depend on the quantum numbers of the sp levels in play.

Based on these results, the dimensionless shift
$\varepsilon_k(N)=\left[\epsilon_k(N)-\epsilon_k(0)\right]/D$
for $k=0,+,-$ is given by
\beq
\varepsilon_0(N)=n_0U\  ,\quad \varepsilon_+(N)=\varepsilon_-(N)=n_0W \  ,
\label{en1}
\eeq
where
$n_k=N_k/(2j_k+1)$ is the occupation number of level $k$,
$U=u(2j_0+1)/D$, and $W=w(2j_0+1)/D$.
It is readily verified that if $fp_FM/\pi^2\sim  1$,
where $p_F=\sqrt{2M\epsilon_F}$ and $\epsilon_F$ is the Fermi energy,
then the integral (\ref{mel1}) has a value
$u\simeq \epsilon_F/A$ and therefore $|U|\sim 1$,
since $D\sim \epsilon_F/A^{2/3}$ in spherical nuclei,

According to Eqs.~(\ref{en1}), the distance $\epsilon_+(N)-\epsilon_-(N)$
remains invariant when $N$ increases.  On the other hand, the
difference $d_+(N)= \varepsilon_+(N)-\varepsilon_0(N)$ decreases
with $N$ when $U>W>0$, as does the distance $d_-(N)=
\varepsilon_0(N)-\varepsilon_-(N)$ in the opposite case, $U<W<0$.

Now let us determine what can happen when the functions $d_{\pm}(N)$
change their signs before the sp level $0$ is completely filled.
We first examine the case $U<W<0$.  According to Eqs.~(\ref{en1}),
the sign of $d_-(N)$ changes at $n_{0c}= 1/(W-U)$, which requires
$W-U$ to be greater than 1 to meet the restriction $n<1$.  The
usual Hartree-Fock (HF) scenario prescribes that for
$n_0>n_{0c}$, quasiparticles must leave the occupied level
$-$ and resettle into the unfilled level $0$.  Further, when
the dependence $\epsilon_{\lambda}=\epsilon_\lambda (n) $ from
Eqs.~(\ref{en1}) is brought into the picture, this effect
is seen to promote the HF rearrangement.

In the opposite case, $U>W>0$, the function $d_+(N)$
changes sign at $n_{0c}=1/(U-W)$, implying  $U-W>1$.
In order to satisfy this inequality, the repulsive part
of the interaction $f$ has to be sufficiently large, or else
the scale $D$ must be rather small.  At $n_0>n_{0c}$,
the HF scenario requires the quasiparticles to leave the unfilled
level $0$ and move into the empty level $+$.  Were this
scenario the correct one, the rearranged sp energies would
obey the equations $\epsilon_0(N)=\epsilon_0(N_c)+\delta N_c(w-u)$
and $\epsilon_+(N)=\epsilon_0(N_c) +\delta N_c(u-w) $,
where $\delta N_c$ is the number of quasiparticles shifted from
level $0$ to level $+$.  The $\delta$ term in each of these equations
arises due to the feedback of the immigrating quasiparticles.
Upon subtracting one equation from the other, we find that
$\epsilon_+(N)-\epsilon_0(N)>0$ for
any $\delta N_c>0$, which says that the level $+$ lies above
rather than below the level $0$.  We thus arrive at a
contradiction that excludes the HF scenario in the case $U>W>0$.

Under these conditions, a new ground state must form, denoted henceforth
by $M$.  As will now be shown, in the state $M$ both of the levels
$0$ and $+$ are partially occupied.  Solution of the problem
for this case reduces to finding the minimum of the
relevant energy functional
\beq
E=\epsilon_0(0)N_0+\epsilon_+(0)N_++{1\over 2}\left[u(N^2_0+N^2_+)
+2wN_0N_+\right]
\label{energy}
\eeq
with $N_k=\sum_m n_{km}$,
through the variational conditions
\beq
{\delta E\over\delta n_{0m}}={\delta E\over\delta n_{+m_1}}=\mu  \,, \qquad
\forall m, m_1 \,,
\label{var}
\eeq
where $\mu$ is the chemical potential.  Such a condition for
characterization of a rearranged ground state first appeared in
Refs.~\onlinecite{ks}, where homogeneous Fermi systems were
addressed, without attention to degeneracy of sp levels.
Eqs.~(\ref{var})
are conveniently rewritten as a conditions for the coincidence of
the sp energies $\epsilon_0$ and $\epsilon_+$,
\begin{eqnarray}
\epsilon_0(N)&=&\epsilon_0(0)+N_0u+N_+w=\mu \,, \nonumber\\
\epsilon_+(N)&=&\epsilon_+(0)+N_0w+N_+u=\mu  \,,
\label{en3}
\end{eqnarray}
which, at $N>N_c=(2j_0+1)/(U-W)$, yield
\beq
{N_0\over N}={1\over 2}\left(1+{N_c\over N}\right)\  , \quad
{N_+\over N}={1\over 2}\left(1-{N_c\over N}\right)  \  .
\label{crit}
\eeq

%%%%%%%%%%%%%%%%%%%%%%%%%%%%%%%%%%%%%%%%%%%%%%%%%%%%%%%%%%%%%%%%
%%%%%               Figure:g0
%%%%%%%%%%%%%%%%%%%%%%%%%%%%%%%%%%%%%%%%%%%%%%%%%%%%%%%%%%%%%%%%%
\begin{figure}[ht]
\centering
\hbox{
\includegraphics[width=0.23\textwidth,height=0.51\textwidth]{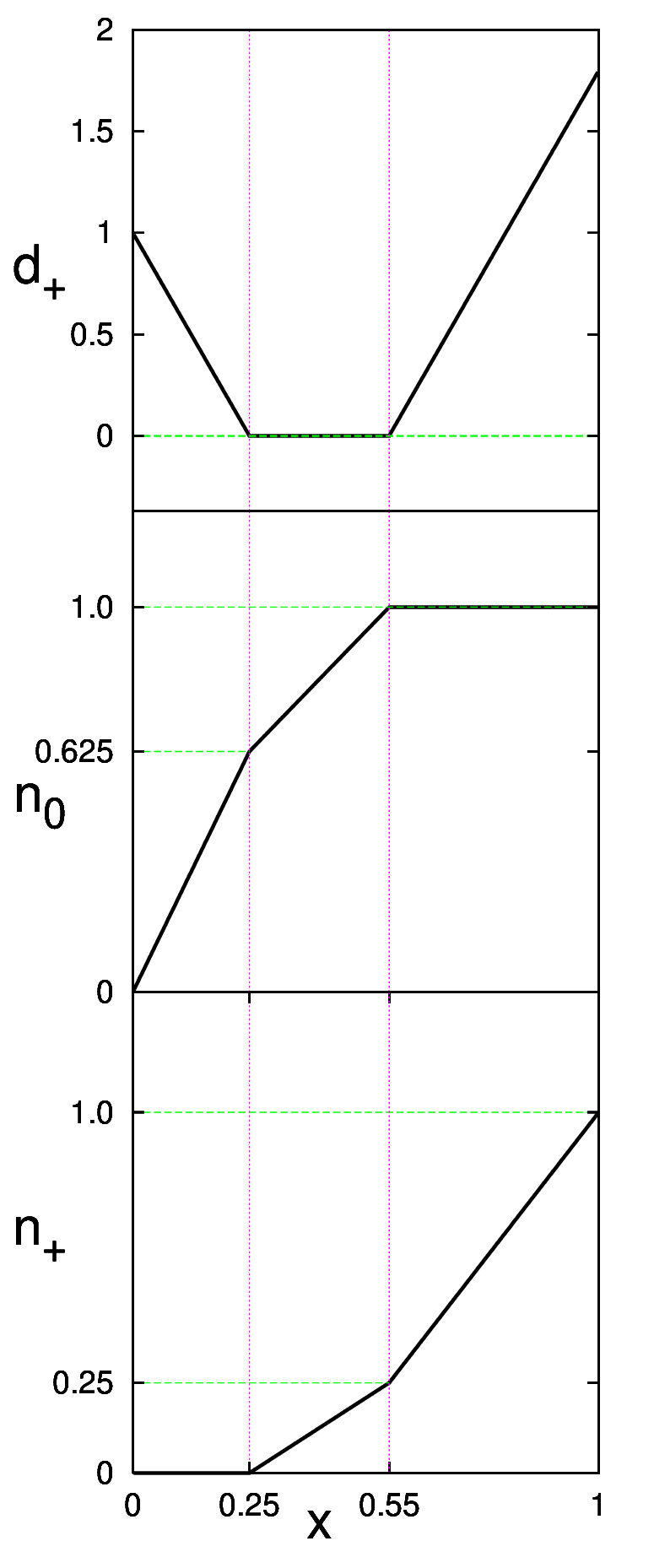}
\includegraphics[width=0.23\textwidth,height=0.51\textwidth]{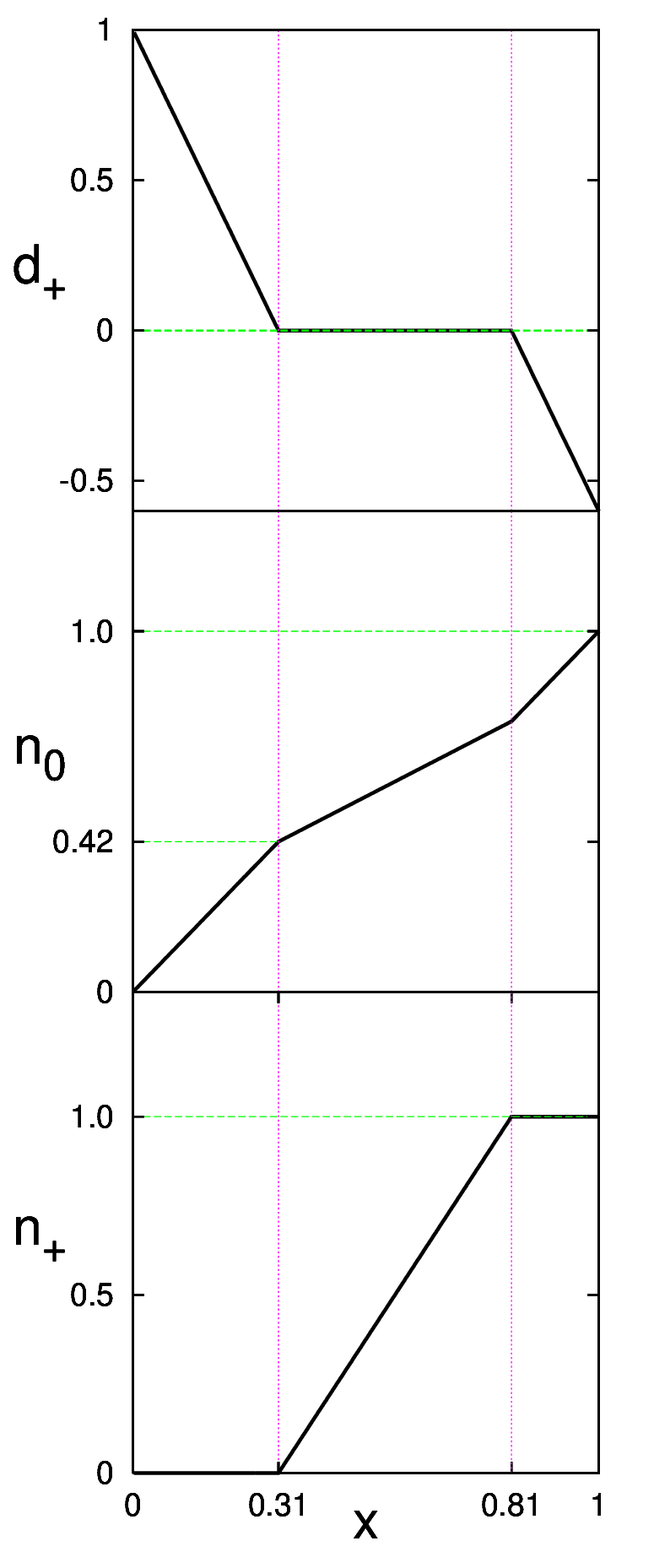}}
\caption{Top panels: Dimensionless distance $d_+=(\epsilon_
+-\epsilon_0)/D$ between levels $+$ and $0$ as a function
of the ratio $x=N/(2j_0+2j_++2)$.  Middle and bottom panels:
Occupation numbers $n_k$ for levels $0$ and $+$.  Input parameters:
$U=4.0, W=2.4$. For the left column, the ratio
$r \equiv (2j_0+1)/(2j_++1)=2/3$; for the right,
$r=3.0$.}
\label{fig:g0}
\end{figure}
%%%%%%%%%%%%%%%%%%%%%%%%%%%%%%%%%%%%%%%%%%%%%%%%%%%%%%%%%%%%%%%%

Results from numerical calculations are plotted in Fig.~\ref{fig:g0},
which consists of two columns, each made up of three plots.
The uppermost panels show the dimensionless ratio $d_+(N)=
(\epsilon_+(N)-\epsilon_0(N))/D$. The middle and lower panels
give, respectively, the occupation numbers $n_0$ and $n_+$.
We observe that there are three different regimes: in two of
them there exist well-defined sp excitations, and $d_+\neq 0$,
and in the third, the energies of the levels 0 and + coincide at
zero.  Passage through the three regimes can be regarded as a
second-order phase transition, with the occupation number $n_+$
treated as an order parameter.

Inserting the above results into Eq.~(\ref{energy}), we find
\beq
E_M-E_{HF}(N_0{=}0,N_+{=}N)=-(u{-}w)(N{-}N_c)^2/4<0 \ ,
\label{gain2}
\eeq
thereby verifying that the $M$ state, having occupation numbers
$0<n<1$ for both of the levels 0 and +, has lower energy than any
HF state.  Significantly, the difference (\ref{gain2}) is of the same
order as a typical shell correction $\delta E_s$ in heavy magic
nuclei.  In such systems, the chemical potential $\mu$ lies in
the large gap between upper filled and lower unoccupied sp
levels, while in the case of merging levels, $\mu$ is located at the
place where the density of states attains its maximum.

The sp levels remain merged until one of them is completely filled.
If the level $0$ fills first, as in the left column
of Fig.~\ref{fig:g0}, then under further increase of $N$,
quasiparticles fill the level $+$, signaling that the distance
$d_+(N)$ again becomes positive.  This behavior resembles the
repulsion of two levels of the {\it same symmetry} in quantum
mechanics, although here one deals with sp levels of
{\it different symmetry}.  In the opposite case where level
$+$ becomes fully occupied before level 0, as in the right
column, the distance $d_+(N)$ becomes negative, and the two
levels just cross each other at this point.

In the nuclear many-body problem, both types of sp level degeneracy
-- either initially present or arising in the scenario described
above -- are lifted when pairing correlations are explicitly
involved.  The role of  $D_{\rm min}$ is played by the pairing gap
$\Delta$ in the spectrum of sp excitations.\cite{belyaev}
To illustrate this situation, we make BCS calculations in the
above two-level model, under the assurance that realistic
pairing forces are weak enough that the gap value remains
smaller than the distance between neighboring sp levels
in magic nuclei.

\begin{figure}[ht]
\centering\hbox{
\includegraphics[width=0.23\textwidth]{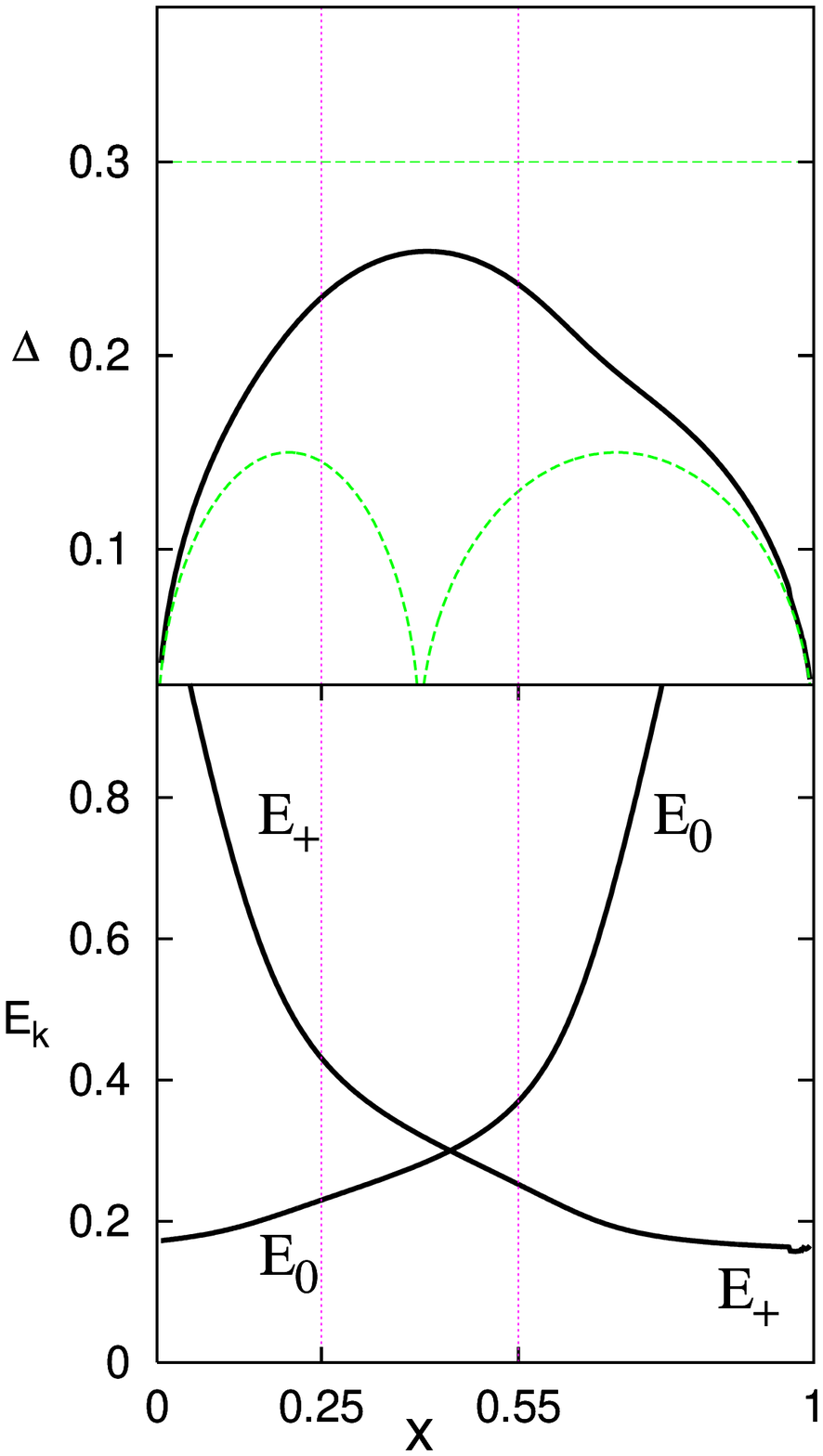}
\includegraphics[width=0.23\textwidth]{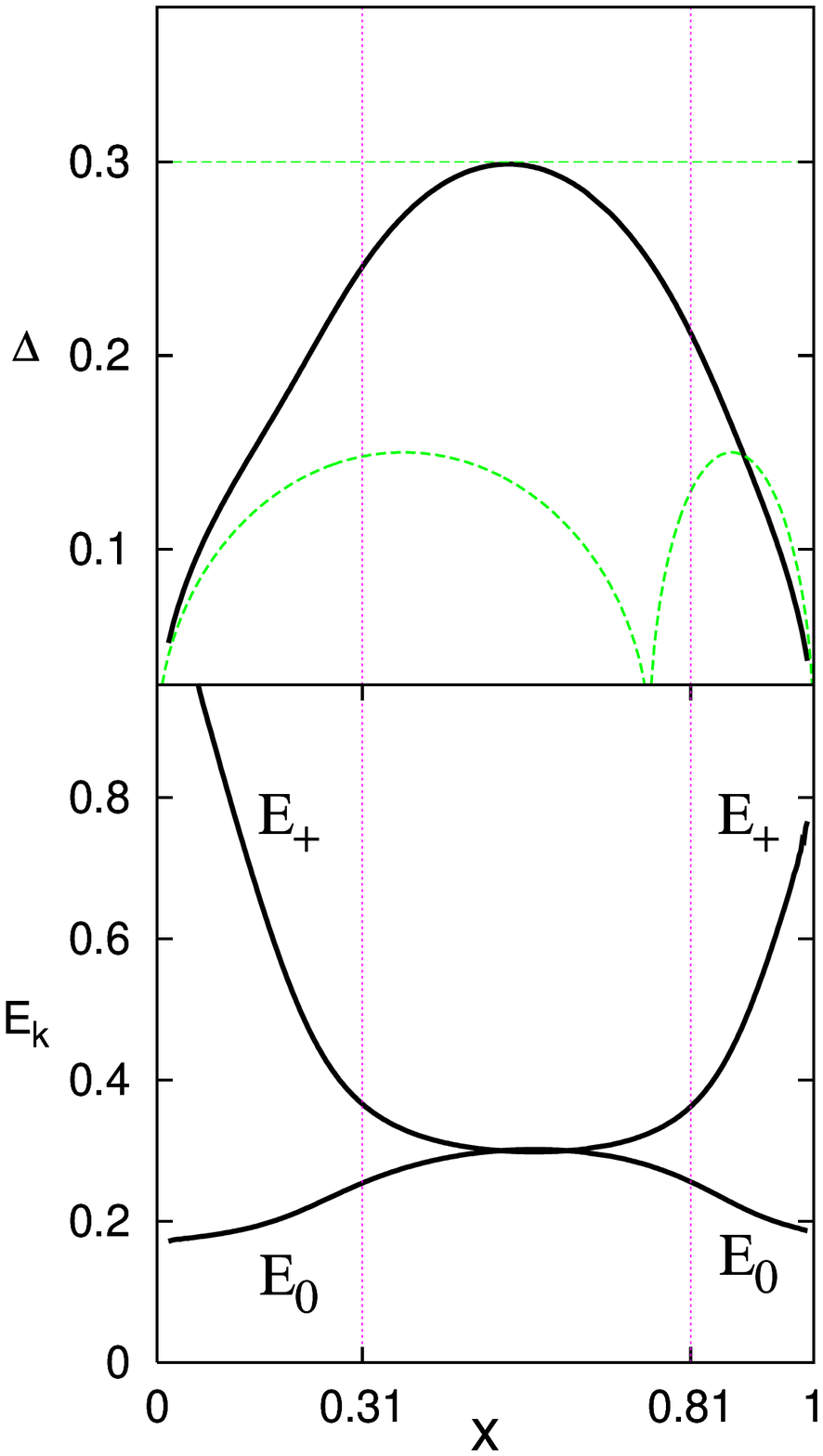}
}
\caption{Top panels: Pairing gap $\Delta$ (in units of $D$)
plotted versus $x=N/(2j_0 + 2j_+ + 2)$, both accounting for the
shrinkage of the interlevel distance (solid line) and neglecting
it (dashed line).  Bottom panels: Energies of Bogoliubov quasiparticles
$E_k=\Delta/2\left[n_k(1-n_k)\right]^{1/2}$.  Pairing constant:
$g= 0.3$.  Other input parameters
for both columns
are the same as in
Fig.~\ref{fig:g0}.}
\label{fig:g03}
\end{figure}

This two-level BCS problem is set up and
solved as follows.  First we rewrite the BCS gap equation as
\beq
\Delta= gD\left(\sqrt{n_0(1-n_0)}+\sqrt{n_+(1-n_+)} \right)\  .
\label{bcs1}
\eeq
In so doing we have followed precedent by introducing
a common dimensionless pairing matrix element
$g=(2j+1)\lambda \epsilon_F/AD$, $\lambda$ being a dimensionless
pairing constant.
A straightforward derivation, based on the BCS identity
$4n_k(1-n_k)=\Delta^2/(\epsilon_k^2+\Delta^2)$, the definition
$\epsilon_{\lambda}=\delta E/\delta n_{\lambda}$ with $E$
given by Eq.~(\ref{energy}), and subtraction of one of Eqs.~(6)
from the other, leads to the key relation
\beq
1+(U-W)(N_0-N_+)={\Delta\over D}\left[R(n_+)-R(n_0)\right]\,,
\label{dpair2}
\eeq
where $R(n_k) = {\rm sgn}(1-2n_k) \sqrt{1/\left[4n_k(1-n_k)\right]-1}$.
and (\ref{dpair2}), together with the obvious equalities $N_k=n_k(2j_k+1)$,
form a closed system determining the occupation numbers ${n_0, n_+}$
and the gap value $\Delta$.  This set of equations must be solved
numerically; some results are given in Fig.~\ref{fig:g03}.  The
inclusion of pairing correlations does indeed lift the degeneracy
of the sp levels.  However, the value of lowest of the energies
$E_k$ of the Bogoliubov quasiparticles remains markedly less
than $D$.

It is instructive to compare the structure of the pairing gap
$\Delta$ in two cases: when the above shrinkage of the
distance between the sp levels $+$ and $0$ is taken into
account, and when it is not.  In the latter case,
$\Delta\sim [n_{\lambda}(1-n_{\lambda})]^{1/2}$
shows two humps with a dip in between.\cite{belyaev}
As seen in Fig.~\ref{fig:g03}, the shrinkage
effect fills in the dip.  This increases the
part of the ground-state energy associated with pairing
correlations.

%%%%%%%%%%%%%%%%%%%%%%%%%%%%%%%%%%%%%%%%%%%%%%%%%%%%%%%%%%%%%%%%%
%%%%%               Figure:g03
%%%%%%%%%%%%%%%%%%%%%%%%%%%%%%%%%%%%%%%%%%%%%%%%%%%%%%%%%%%%%%%%%
%%%%%%%%%%%%%%%%%%%%%%%%%%%%%%%%%%%%%%%%%%%%%%%%%%%%%%%%%%%%%%%%

Let us now address the case $u>w>0$ without pairing,
existing for example in atoms and quantum dots.  In this case, a pair
of particles added to any sp level with $l\neq 0$ always have total
angular momentum $J\neq 0$ (Hund's rule), in principle destroying spherical
symmetry and lifting the $m$-degeneracy of the sp energies
$\epsilon_{km}$. This gives rise to spreading of the levels,
the magnitudes of which are proportional to $u$ for the
level $0$, and $w$ for the level $+$.  If the interaction
function $f$ has long-range character, we have $u/w\gg 1 $,
and hence the spread of level $0$ is much larger than that of
level $+$.  For the density $\delta\rho$ associated with the
added quasiparticles, we may write $\delta \rho({\bf r}) =
R^2_{n_0l_0}(r)\sum_m| \Phi^2_{j_0l_0m}({\bf n})|^2n_{0m}$,
which is applicable at least until the crossing of relevant
orbitals begins.  Upon inserting this formula into the relation
$\delta\Sigma=(f\delta\rho)$, it is found that the spread
does not affect the evolution of the centers of gravity
$\epsilon_k^{\rm o}= \sum_m \epsilon_{km}/(2j_k+1)$ of the levels,
since the isotropic part of $\delta \rho$ has the same
form $\delta_0\rho=NR^2_{n_0l_0}(r)/4\pi$ as if the degeneracy
of the sp level were still in effect.  This circumstance is
especially important at the stage when the two families of
sp levels begin to cross each other.  Since at $u>w>0$ the
center of the gravity of the level $+$ gets stuck close
to the Fermi surface, our results provide a simple
mechanism for pinning of the narrow bands in solids
to the Fermi surface.

To exemplify this point, let us consider a model where the
sp spectrum in local-density approximation (LDA) is exhausted
by (i) a wide band, which disperses through the Fermi surface,
and (ii) a narrow one, placed below the Fermi surface at a
distance $D_n$.  We assume that only the diagonal matrix element
$f_{nn}$ of the interaction function $f$ referring to the
narrow band is significant, the others being negligible.
The shift $\delta\epsilon_n$ in the location of the narrow band due
to switching on the intraband interactions is given by a formula
analogous to Eq.~(3), namely $\delta\epsilon_n =f_{nn}\rho_n$,
where $\rho_n$ is the density of the band.  If the correction
$\delta\epsilon_n$ exceeds the distance $D_n$ then the HF
scenario calls for the narrow band to be completely emptied;
but then the shift $\delta\epsilon_n$ must vanish.
To eliminate this inconsistency, only a fraction of the particles
leave the narrow band, in just the right proportion to equalize
the chemical potentials of the two bands.  The feedback mechanism
we have described positions the narrow band exactly at the Fermi
surface, resolving a long-standing problem with the LDA
scheme.

In atoms, remnants of an accidental degeneracy of the Coulomb
problem persist in the formation of electronic shells for
which the distance between sp levels with different orbital
momenta $l$ is rather small.  Recalling that matrix elements
of the electron-electron interaction are quite sensitive
to the $l$ value, mergence of definite sp levels cannot
be excluded. To elucidate this situation, one needs to
analyze the energy functional $E=\sum\epsilon_k(0)n_{km}+
{1\over 2}\sum u_{km,k_1m_1}n_{km}n_{k_1m_1}$, wherein
the interaction matrix $u_{km,k_1m_1}$ replaces the matrix
element (\ref{mel1}) and summation occurs over some states
of the last unfilled shell.  Results from numerical
studies of the variational equations
generalizing Eqs.~(\ref{var}),
$ \mu=\epsilon_k(0)+\sum u_{km,k_1m_1}n_{k_1m_1}$,
will be given elsewhere.

The new many-body effect uncovered in the foregoing analysis
resembles a previously studied phenomenon, called
fermion condensation, which involves wholesale mergence of
sp levels in homogeneous Fermi fluids.\cite{ks,physrep}
In any conventional homogeneous Fermi liquid, e.g.,
liquid $^3$He, the momentum ${\bf p}$ of an added particle
can be associated with a certain quasiparticle.  Similarly,
in most spherical odd-$A$ nuclei, the total angular momentum
$J$ in the ground state is carried by an odd quasiparticle.
In atomic physics, the electronic configuration of ions of elements
belonging to the principal groups of the periodic table repeats
that of preceding atoms.  From the microscopic perspective,
in all such ``open-shell'' systems conforming to standard FL
theory, the single quasiparticle term $a^+_{\lambda}\Psi_0$
assumes a special role in the ground-state wave function,
where $\Psi_0$ represents the ground state of a parent system.
By contrast, in the case of merging of sp levels, the ground-state
features a multitude of quasiparticle terms and therefore exhibits
a more complicated, yet more balanced character ---
as in the comparison of a chorus with a dominant soloist.
This implication of our analysis offers a qualitative
explanation of the fact that the chemical properties of
rare-earth elements differ little, in spite of marked
variation in atomic numbers. Such an explanation
may or may not be at variance with the textbook
argument\cite{landau} that the relative squeezing of
$f$ and $d$ orbitals is responsible for the remarkable
similarity.

In spite of  evident commonalities, there is a crucial
difference between the conditions for the ``level-mergence''
phenomenon in homogeneous Fermi liquids and in finite Fermi
systems with the degenerate sp levels.  In the former, the
presence of a significant velocity-dependent component in
the interaction function $f$ is needed to promote fermion
condensation, while in the latter, sp levels can merge
even if $f$ is momentum-independent. The reason for
this difference is simple: in the homogeneous case, the
matrix elements $u$ and $w$ are equal to each other,
implying zero energy gain due to the rearrangement when
velocity-dependent forces are absent. We point out that
the study of level mergence in finite systems has the
advantage of transparency,
in that (i) it is free of
the complicated issue of damping sp excitations, and (ii)
it gives access to the precursor stage of the effect.

Our exploration of the mergence of single-particle
levels in a finite Fermi system with degeneracy has revealed
a phenomenon that entails the disappearance of well-defined
low-lying single-particle excitations, with important
implications in diverse physical settings.

We thank K.~Kikoin, Z.~Nussinov, E.~Saperstein, and V.~Yakovenko
for fruitful discussions. This research was supported by
Grant No.~NS-8756.2006.2 from the Russian Ministry of
Education and Science.


\begin{thebibliography} {99}
\bibitem{migdal} A.~B.~Migdal, {\it Theory of Finite Fermi Systems and
Applications to Atomic Nuclei} (Wiley, New York, 1967).
\bibitem{bohr} A.~Bohr and B.~Mottelson, {\it Nuclear Structure},
Vol.~1 (W.~A.~Benjamin, inc., New York, 1969).
\bibitem{shoenberg} D.~Shoenberg, {\it Magnetic Oscillations}
(Cambridge University Press, Cambridge 1984).
\bibitem{schuck} P.~Ring and P.~Schuck, {\it The Nuclear Many-Body Problem}
(Springer-Verlag, Berlin, 1980).
\bibitem{ks} V.~A.~Khodel and V.~R.~Shaginyan, JETP Letters {\bf 51},
553 (1990); Condensed Matter Theories {\bf 12}, 221 (1997).
\bibitem{physrep}  V.~A.~Khodel, V.~V.~Khodel, and V.~R.~Shaginyan,
Phys.~Rep.~{\bf 249}, 1 (1994); V.~A.~Khodel, M.~V.~Zverev, and
V.~M.~Yakovenko, Phys.~Rev.~Lett.~{\bf 95}, 236402 (2005).
\bibitem{belyaev} S.~T.~Belyaev, Mat.~Fys.~Medd.~Dan.~{\bf 31}, No.~11 (1959).
%~Vid.~Selsk.
\bibitem{landau}
L. D. Landau and E. M. Lifshitz, {\it Quantum mechanics,
Non-relativistic theory} (Pergamon, New York, 1965), p.~290.
\end{thebibliography}
\end{document}